\documentclass[a4paper,usenatbib,onecolumn]{mn2e}
\usepackage{newtxtext,newtxmath}
\usepackage{lipsum}
\usepackage[T1]{fontenc}
\usepackage{ae,aecompl}
\usepackage{hyperref}
\usepackage{graphicx}   % Including figure files                                                           
\usepackage{amsmath}    % Advanced maths commands                                                          
\usepackage{amssymb}    % Extra maths symbols                                                              
\usepackage{bm} % Extra maths symbols                                                                      
\usepackage{caption}
\usepackage{multirow}
\usepackage{multicol}
\usepackage{cuted}
\usepackage{mathtools}
\usepackage{longtable}

\def \apjs {{\rm {ApJS}}}

\def \aap {{\rm {A\&A}}}

\def \apjl{\rm {ApJL}}

 \title{Minimizing the bias in exoplanet detection - application to radial 
velocities of LHS 1140}
\author[F. Feng et al.]
{F. Feng$^{1}$\thanks{E-mail: f.feng@herts.ac.uk or fengfabo@gmail.com}, M. Tuomi$^{1}$, H. R. A.  Jones$^{1}$\\
$^{1}$Centre for Astrophysics Research, University of Hertfordshire, College
Lane, AL10 9AB, Hatfield, UK}
\date{\today}

\begin{document}
\maketitle
\begin{abstract}
  A rocky planet orbiting LHS 1140 with a period of 24.7d has been found based on the discovery of transits in its light and high precision radial velocity data \citep{dittmann17}. This discovery by two independent methods is an observational tour-de-force, however, we find that a conservative analysis of the data gives a different solution. A three planet system is apparent in the radial velocity data based on our diagnosis of stellar activity. We encourage further targeted photometric and radial velocity observations in order to constrain the mini-Neptune and super-Earth mass objects apparently causing the 3.8 and 90 day radial velocity signals. We use our package Agatha (\url{https://phillippro.shinyapps.io/Agatha/}) to provide a comprehensive strategy to disentangle planetary signals from stellar activity in radial velocity data.
\end{abstract}
%%%%%%%%%%%%%%%%%%%%%%%%%%%%%%%%%%%%%%%%%%%%%%%%%%%%%%%%%%%%%%%%%%%%%%%%%%%%
\section{Introduction}\label{sec:introdution}
Over the last few years, a few small planets in the temperate zone of M-dwarfs have been detected and characterized using transit and radial velocity (RV) methods. For example, the Trappist-1 system is composed of seven Earth-sized planets \citep{gillon16} while at least one planet is found to orbit Proxima Centauri \citep{anglada16}. Considering that M dwarfs are relatively light and easily perturbed by planets, they are optimal targets for detecting potentially habitable planets using the radial velocity method.

The RV of a star is not only periodically perturbed by planets but is also modulated by the stellar activity, which changes the brightness and also the spectrum of the star. Hence the noise in RV data is typically correlated in time and in wavelength and should be modeled with appropriate red noise models \citep{baluev13,feng17c}. Without proper noise modeling, the signal is either interpreted as noise by inappropriate red noise models or the noise is interpreted as signal by white noise models \citep{feng16}. To avoid these situations, a so-called ``Goldilocks noise model'' should be selected to avoid false negatives and false positives \citep{feng16}. A Goldilocks model should be able to model both the time-correlated and wavelength-correlated noise in the data. Based on a comparison of various noise models, we suggest the use of the moving average (MA) model in linear combination with noise proxies such as Bisector, strength of CaII-HK lines, and differential RVs defined in \citep{feng17c}. Some of these combined noise models may be equally plausible based on the Bayesian model comparison. If the significance of a signal is found to be sensitive to the choice of these noise models (e.g. the MA model with different numbers of noise proxies), it is indicative that it is unlikely to be related to planet (e.g. \citealt{feng18a}). 

The traditional Lomb-Scargle periodogram \cite{scargle82} is typically used to identify and visualize signals in RV time series, however, it assumes white noise in the data and thus inappropriate to properly find the signal and assess its significance reliably. More advanced red noise periodograms such as ``Bayes factor periodogram'' (BFP) have been developed by \cite{feng17a} to model the red noise and signal in the data simultaneously. A framework of red noise periodograms and model selection methods have been implemented in the {\small Agatha} software \citep{feng17a}. It has been successfully used by \cite{feng17b}, \cite{feng17b}, and \cite{feng18d} to detect small planets in the RV data. In this work we use {\small Agatha} in combination with MCMC methods to analyze the RV data of LHS 1140 which is found to host one planet with a period of 24.7\,d through a combined fit of transit and RV data by \cite{dittmann17} (hereafter D17). In this paper, we first use the BFP to find signals in the RV data for LHS 1140 section \ref{sec:rv}. We then investigate into the nature of the 90\,day signal in the Agatha framework in section \ref{sec:90day}, followed by a test of consistency of signals over time in section \ref{sec:consistency}. Finally we conclude in section \ref{sec:conclusion}.

\section{Analysis of LHS 1140 radial velocities}\label{sec:rv}
\begin{figure*}
  \centering
\includegraphics[scale=0.6]{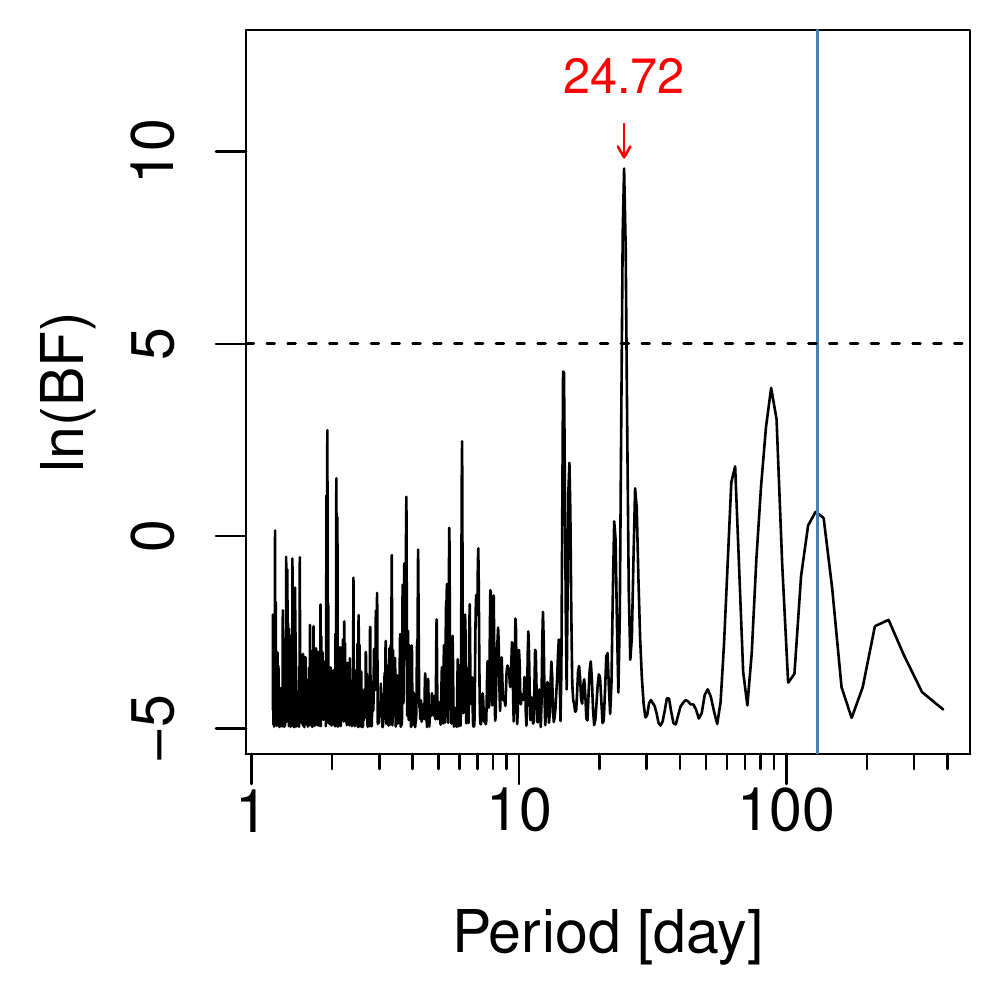}
\includegraphics[scale=0.6]{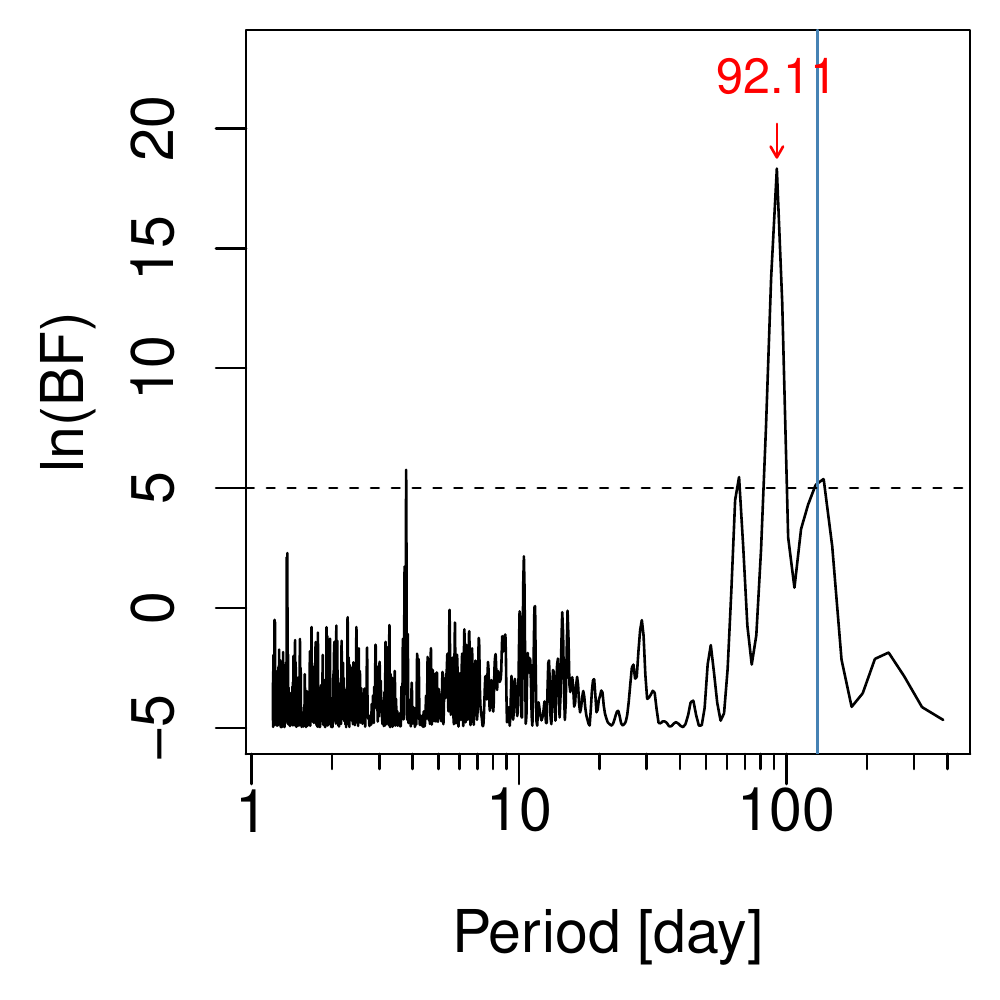}
\includegraphics[scale=0.6]{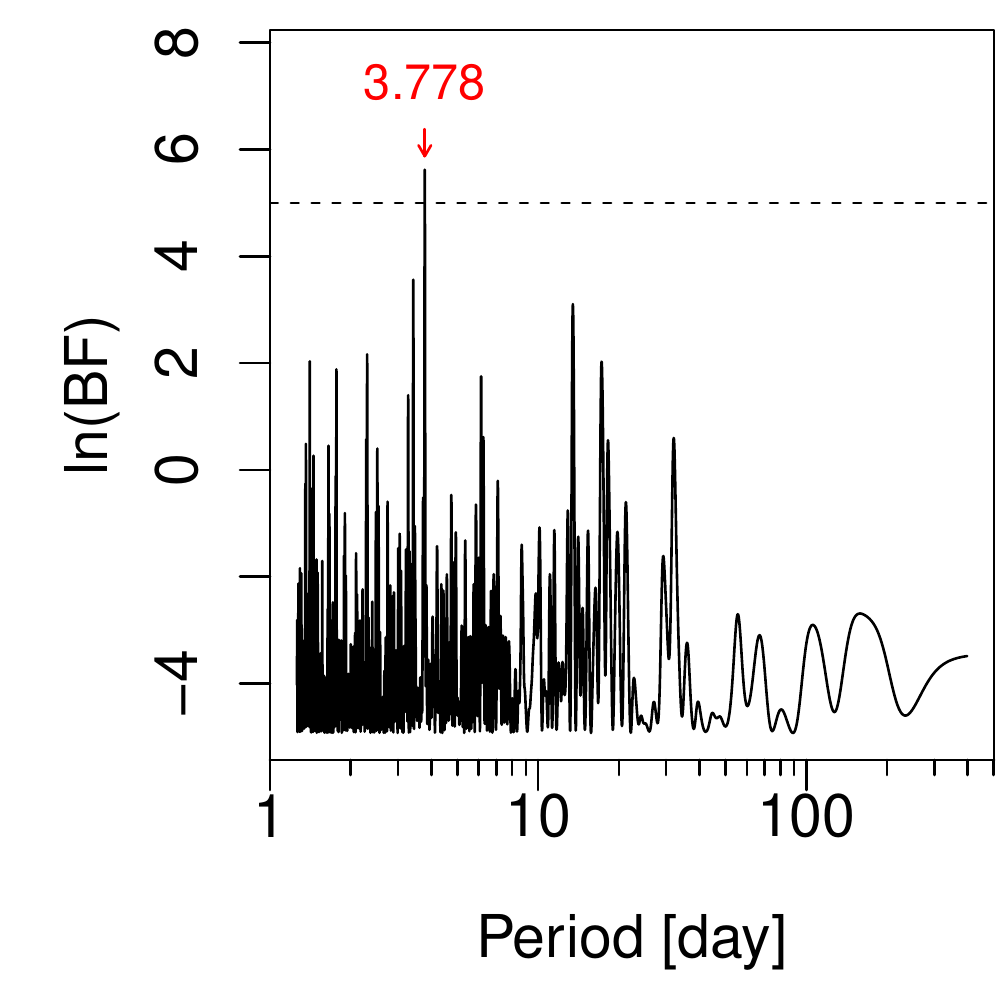}
\includegraphics[scale=0.6]{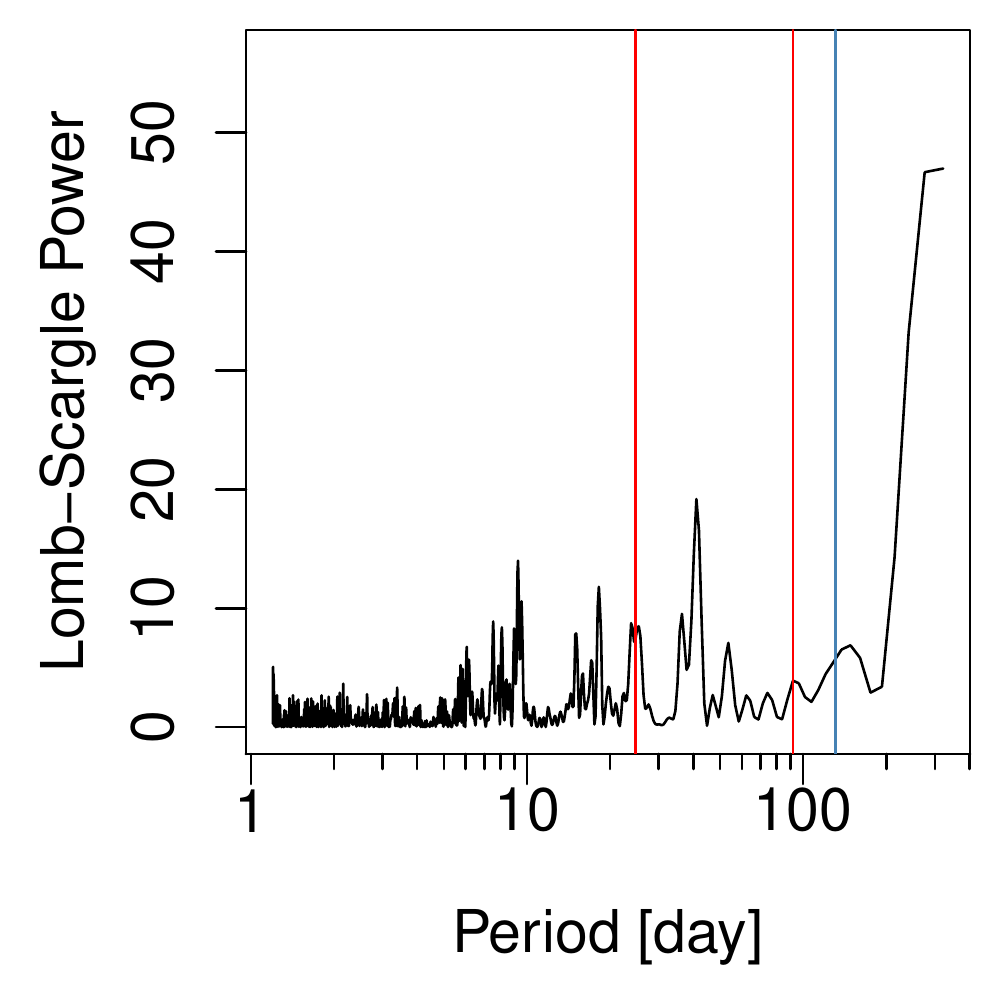}
\caption{BFPs for the radial velocity data (upper left) and for the data after subtracting the circular 24.7\,d signal (upper right) and the 92\,d signal (bottom left) and the Lomb-Scargle periodogram of the window function (right). The rotation period of 131\,d is denoted by the blue lines. The signals are denoted by arrows (upper and bottom left panels) and by red lines (bottom right panel). The Bayes factor threshold of 150 is denoted by the dashed lines. }
\label{fig:BFP_2sig}
\end{figure*}

In D17, the nearby star LHS1140 is found to have a regular photometric transit based on data from several different telescopes and that the same signal is apparent in follow-up radial velocities. In coming to this conclusion, D17 assign a first signal at 130\,d in the photometric data for which they find some evidence for in the radial velocity data as arising from rotation and secondly a 90\,d signal in the radial velocity data as arising from rotation-modulated stellar activity or by time samplings or by both. Their analyses of these signals is based on signal visualization using Lomb-Scargle periodograms. However, such periodograms do not include time-correlated noise. To account for this noise and to explore the nature of this signal, we calculate the BFP\footnote{The code for the calculation is available at \url{https://github.com/phillippro/agatha} and the corresponding application is at \url{https://phillippro.shinyapps.io/Agatha/} and \url{http://agatha.herts.ac.uk/}.} to explore the nature of this signal. In the calculation, we adopt a combination of a linear trend, a constant excess noise, and the first order moving average model (or MA(1)) as the baseline model, which is found to be one of the best noise models for radial velocity data according to the RV-challenge competition \citep{dumusque16b}. To account for wavelength-dependent noise, we also include a linear function of the 3AP2-1 differential RVs \citep{feng17c} based on a comparison of different noise models \citep{feng16}. From the periodograms shown in Fig. \ref{fig:BFP_2sig}, we see a significant signal at a period of about 90\,d which is not found to be strong in the window function. We also see a signal at a period of about 3.8\,d in the residual BFP. If the 90\,d signal was caused by the window function, the 24.7\,d signal must also be false since the power of the latter is stronger than the former in the periodogram of the window function.

Without investigating into the contribution of stellar rotation to the radial velocity variation, D17 have subtracted the signal at the rotation period of 130\,d from the data and claim a strong association of the 90\,d signal and the rotation period. This is not supported by our analysis. Adopting the Gaussian process model used by DT17, we initialize Markov chains at the rotation period with the semi-amplitude given by D17, and draw posterior samples from the posterior. But the chains cannot converge to a stationary posterior distribution at the rotation period, indicating a low posterior around the rotation period.

\begin{figure*}
  \centering
\includegraphics[scale=0.5]{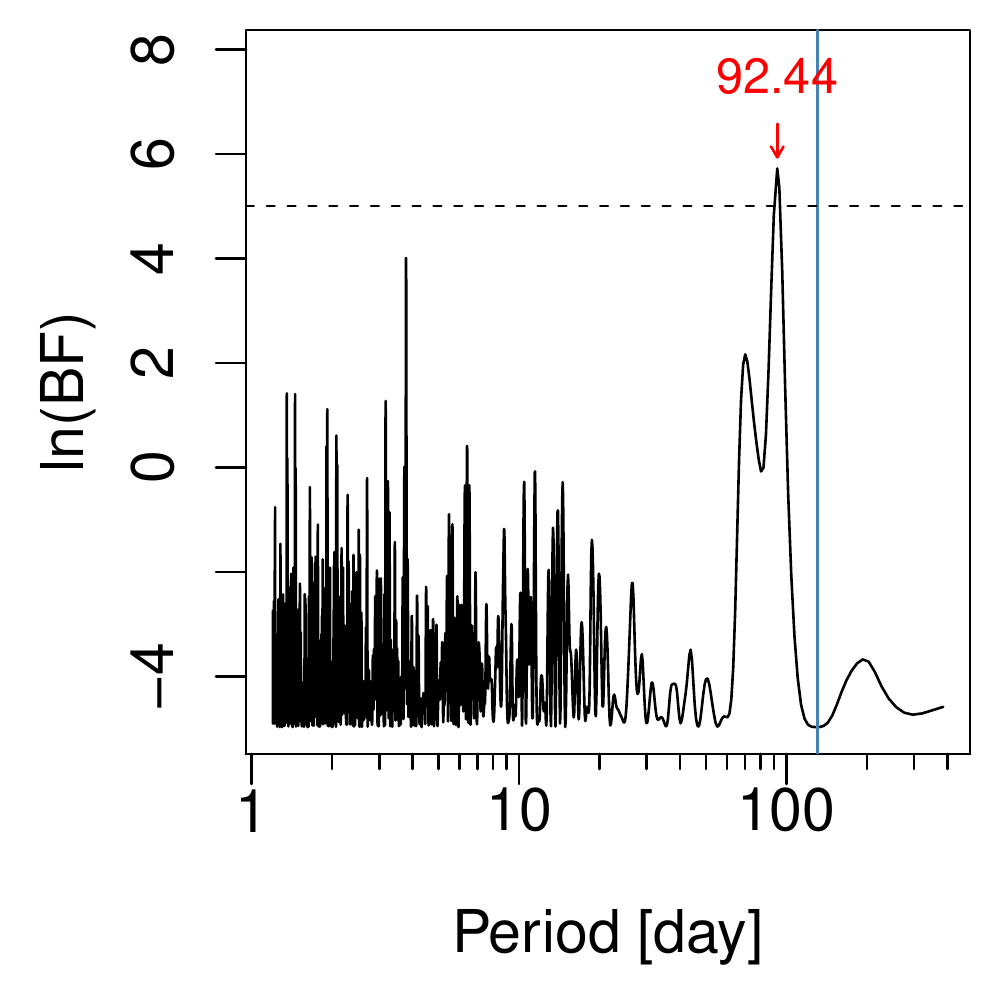}
\includegraphics[scale=0.5]{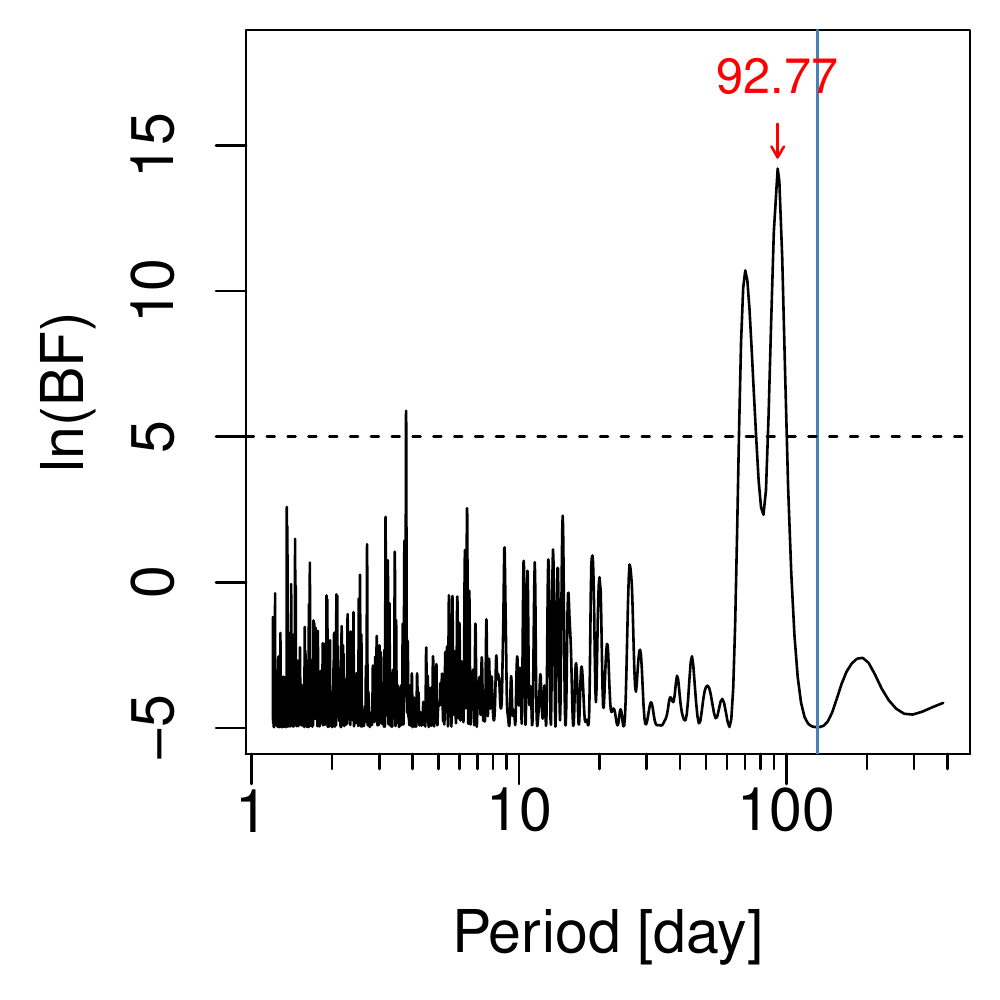}
 \includegraphics[scale=0.5]{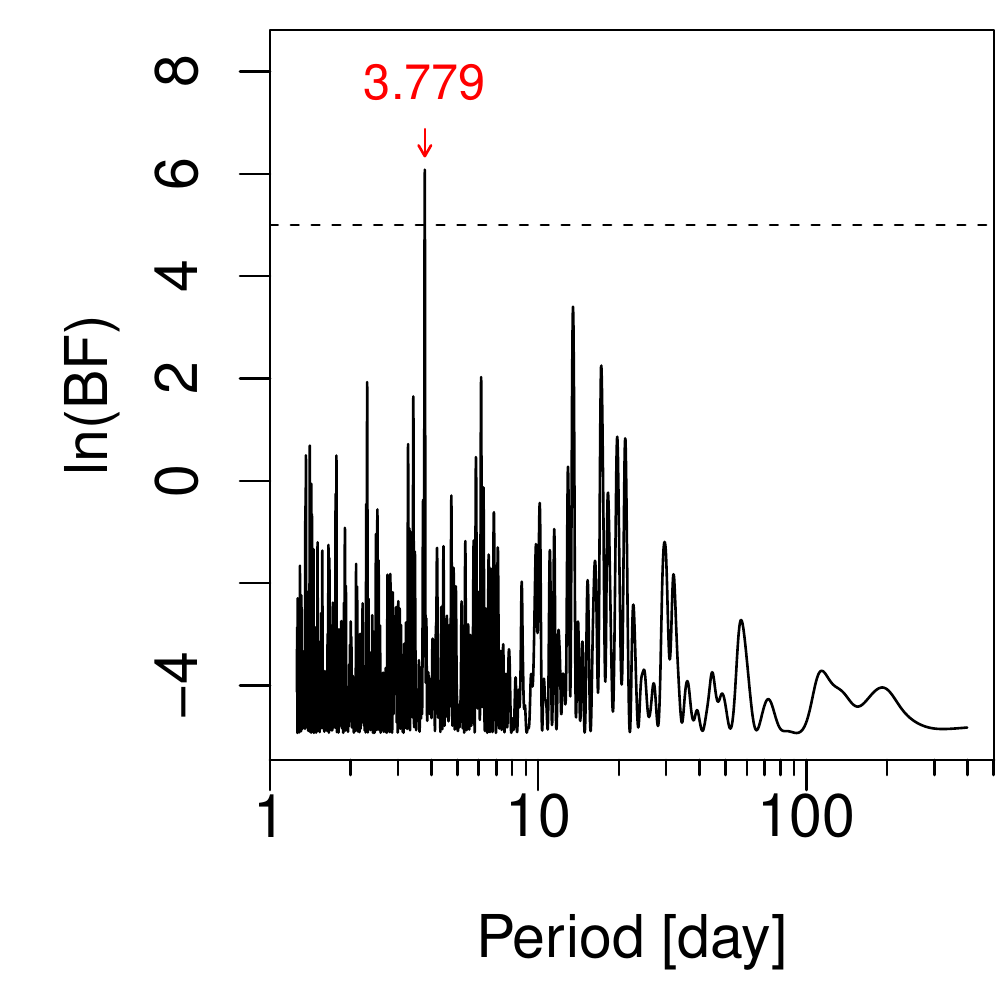}
\caption{ BFPs for RV residuals. The left and middle panels respectively show the impact of subtracting the rotation-induced signal at a period of 130\,d and a period of 24.7d on the BFPs for the MA(1) (left) and white noise (middle) models. The right panel shows the white noise BFP of the data subtracted by the 24.7 and 90\,d signals. }
\label{fig:activityBFP}  
\end{figure*}

We further apply BFP to find the orbital parameters of the 130\,d signal and subtract this signal together with the 24.7\,d signal from the data. When we examine the residuals after subtracting these signals in both the left and middle panels of Fig. \ref{fig:activityBFP} we see the 90 d signal remains significant. Thus we observe the persistency of the 90\,d signal despite being weakened compared with the middle panel of Fig. \ref{fig:BFP_2sig} due to a subtraction of the 130\,d signal. The 90\,d signal is more significant in the white noise BFP because the MA(1) model tends to reduce the significance of time-correlated noise and real signals simultaneously \citep{feng16}. In the BFP shown in the right panel of Fig. \ref{fig:activityBFP}, the 3.8\,d signal is unique and passes the BF threshold. We investigate the signals at periods of $\sim$90\,d and 3.8\,d in the following sections. 
% We still find the signal strongly 
%\vspace{1in}

\section{Investigate into the nature of the 90 day signal}\label{sec:90day}
Since the Gaussian process used by D17 and in this work has already accounted for the activity-induced signals, the persistency of the 90\,d signal strongly support its Keplerian origin. Moreover, we adopt various noise models to analyze the data to test the consistency of this signal to see whether the presence of the signal is dependent on the selected noise model. 

In addition to the Gaussian process model, we define the white noise model as a linear trend combined with a constant excess noise (white noise model). Its combination with the first and second moving average models defines the MA(1) and MA(2) models. Based on the Bayesian method implemented by adaptive MCMC posterior sampling, we find that the 90\,d signal is insensitive to the choice of noise models and has a semi-amplitude of about 5.6\,m/s, comparable with the semi-amplitude of the 24.7\,d signal (see Table 1 of DT17). For all these noise models, the addition of the 90\,d signal increases the maximum likelihood by more than 8 orders of magnitude and increases the Bayes factor\footnote{This Bayes factor is estimated by the Bayesian information criterion (BIC) and is considered a conservative metric of the plausibility of a model based on a comparison of various Bayes factor estimators \citep{feng16}. } by a factor of 2000, strongly favoring a two-planet radial velocity model based on the threshold of 150 given by \cite{kass95} and \cite{feng16}. 

We continue the investigation of the nature of the 90\,d signal by injecting synthetic signals into the residuals after subtracting the 24.7 and 92\,d signals quantified using the Gaussian process model. We inject signals at periods of 3, 5, 12, 17, 59, 109, 130 and 310\,d with semi-amplitudes of 0.4, 0.8, 1.6, 3.2, 5.6, 6.4, 12.8 and eccentricities of 0.0 and 0.2 into the residual. Based on posterior samplings drawn by adaptive MCMC chains with more than 3 million samples for each, we are able to recover signals with periods shorter than 50\,d with semi-amplitude of 3.2\,m/s. We also recover the signals shorter than 130\,d with semi-major amplitude of 5.6\,m/s, which is the semi-amplitude of the 90\,d signal. Moreover, all signals with a semi-amplitude of 12.8\,m/s can be recovered, including the 130\,d signal which is the rotation period modeled by the Gaussian process and injected into the D17 analysis. Thus our detection of the 90\,d signal with a semi-amplitude of 5.6\,m/s is plausible considering that signals with similar periods and amplitudes can be recovered.

\begin{figure}
  \centering 
  \includegraphics[scale=0.45]{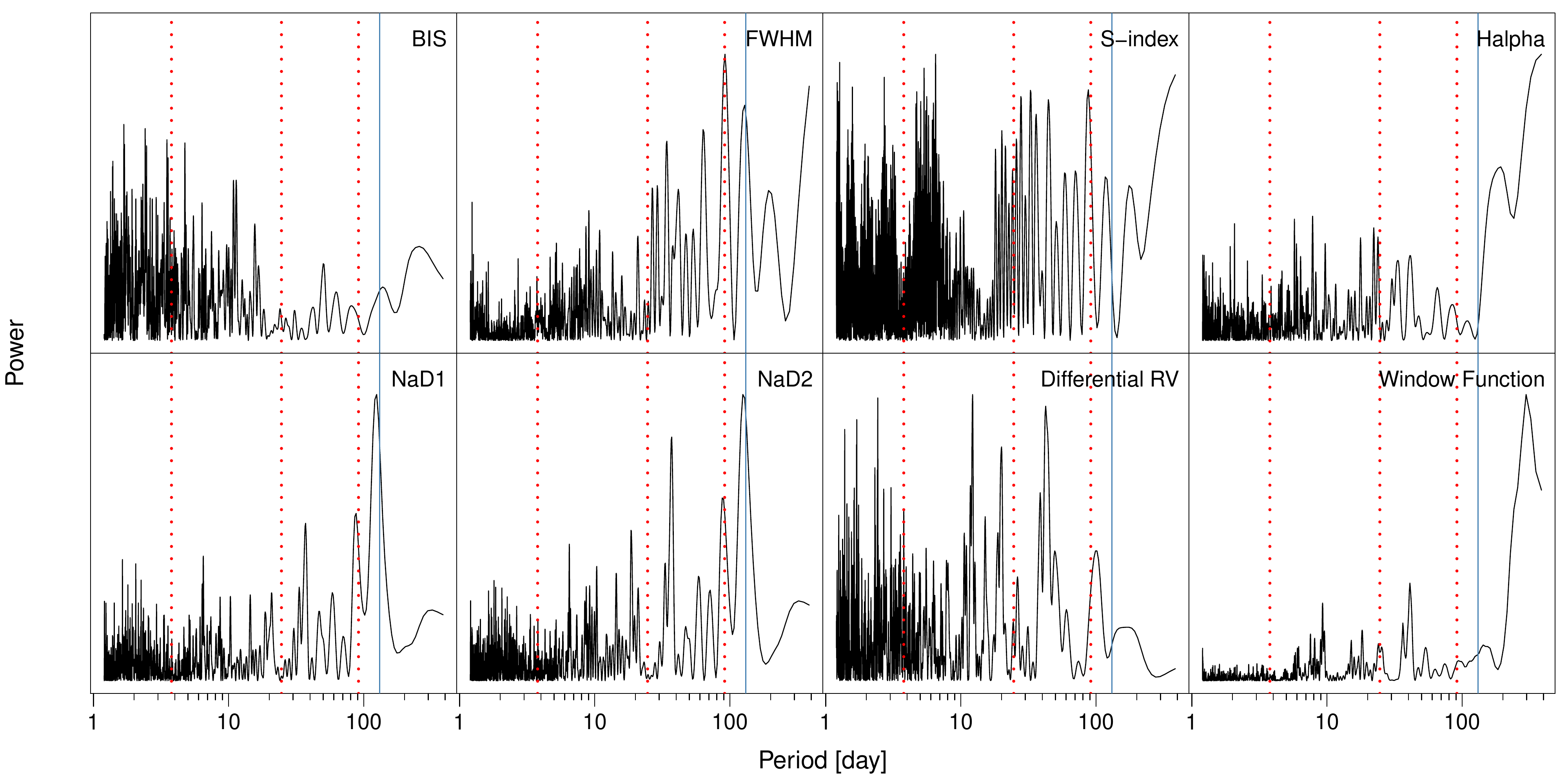}
  \caption{GLSTs of the calcium activity index (S-index), the line bisector span (BIS) and full width half maximum (FWHM) of the spectral lines, the intensity of H-alpha, Sodium lines (NaD1 and NaD2) and differential RVs. The BFPs are calculated based on a white noise model. }
\label{fig:indices}  
\end{figure}

\begin{figure}
  \centering 
\includegraphics[scale=0.8]{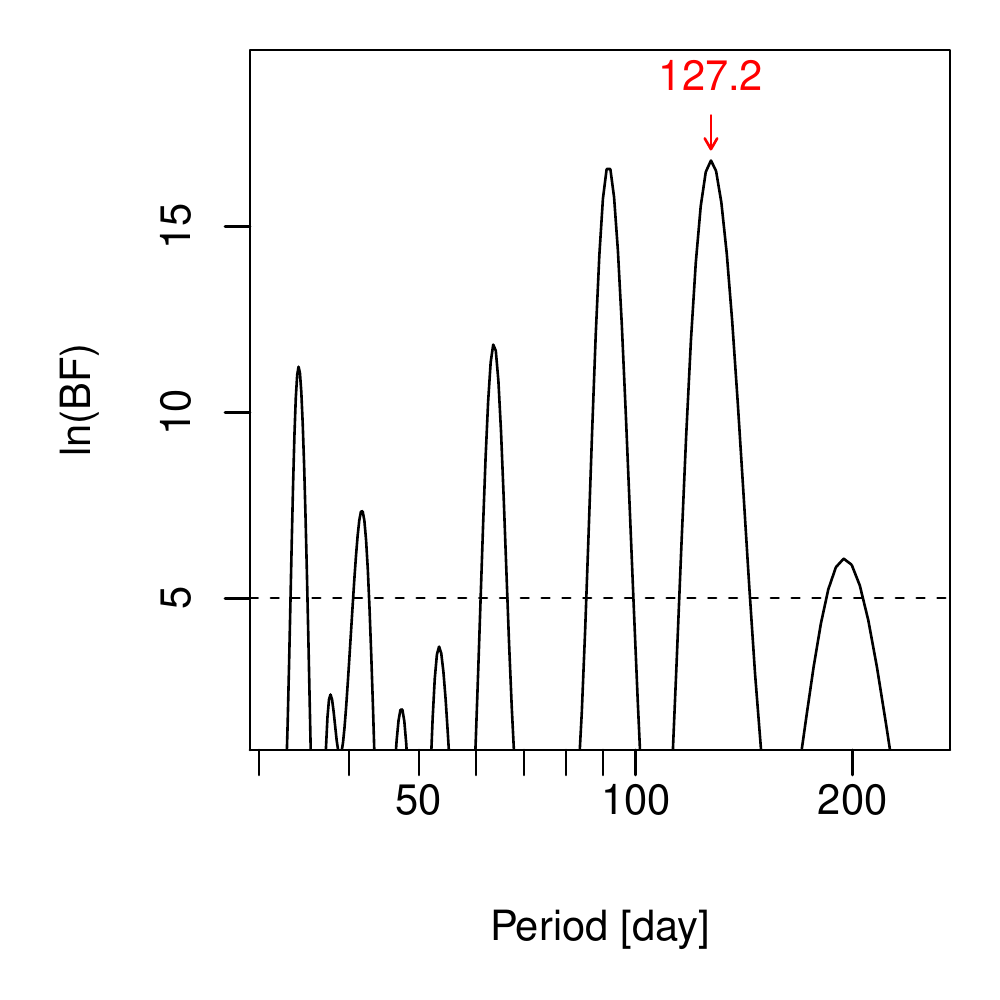}
\includegraphics[scale=0.8]{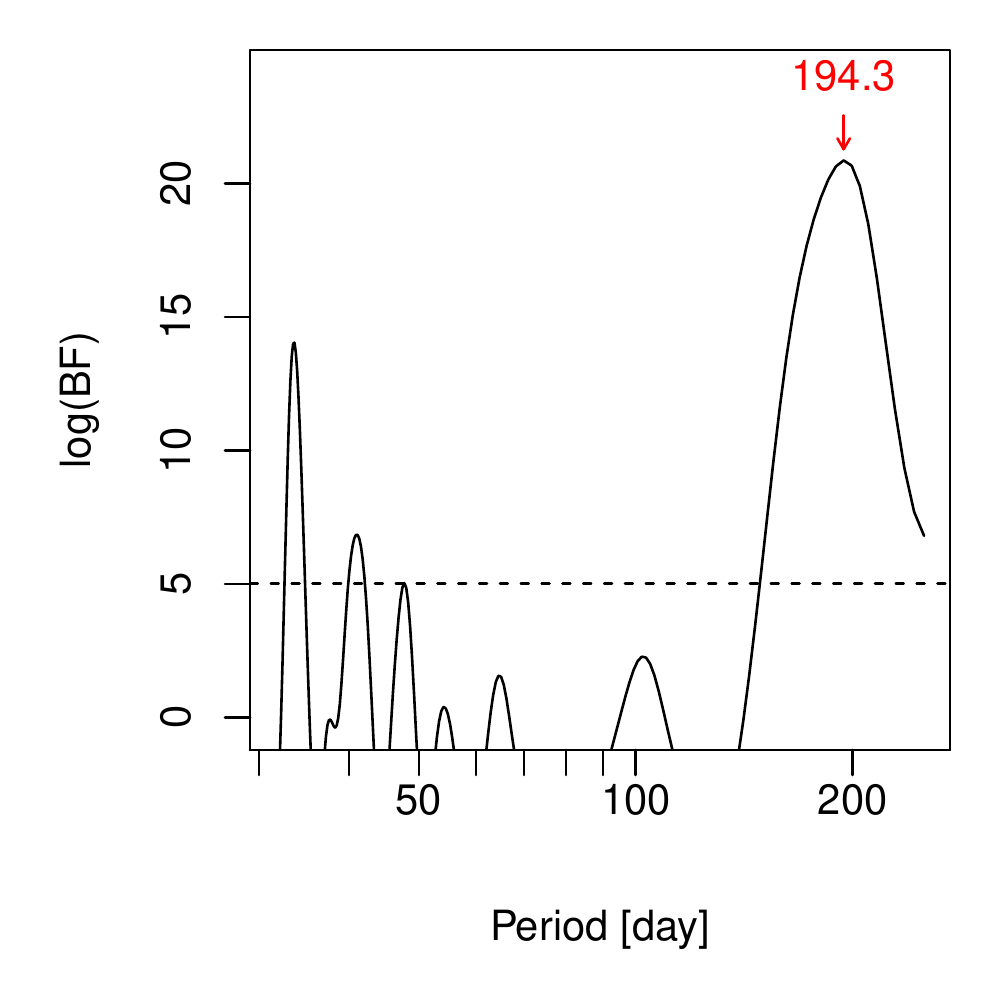}
\caption{Zoomed BFPs of the FWHM (left) and the FWHM after subtraction of the stellar activity signal (right)}
\label{fig:FWHM}  
\end{figure}

To check whether the 90\,d signal is caused by stellar activity, we show the generalized Lomb-Scargle with floating trend (GLST) periodograms \citep{feng17a} of various activity indices together with the signals in Fig. \ref{fig:indices}. We observe that the signals do not overlap with the strong powers in the BFP plots of S-index, BIS and H-alpha. But the long period signals are strong in the GLSTs of FWHM, and NaD1 and NaD2. We find that the periodogram powers around 90\,d and 130\,d are high for the FWHM, NaD1 and NaD2 (measuring the strength of sodium lines). Since such an overlap does not appear in other indices, the FWHM and sodium lines are probably particularly sensitive to stellar activity. Based on further analysis (see Figure \ref{fig:FWHM}), we see a strong connection between the 90\,d signal and the activity in the FWHM and sodium line variation since the 90\,d signal disappears after the subtraction of the activity signal. But such a strong connection is not found in the radial velocity data (see Fig. \ref{fig:activityBFP}). Moreover, the 90\,d signal is much more significant than the activity signal in the radial velocity data (see the top right panel of Fig. \ref{fig:BFP_2sig}). Therefore we confirm the 90\,d signal as a planetary candidate. Furthermore, the 3.8\,d signal is significant in the white noise BFP of the residuals after subtracting the first two signals (see Fig. \ref{fig:activityBFP}), and is visible in the red noise BFP (see Fig. \ref{fig:BFP_2sig}). 

\section{Consistency of the 3.8 day signal over time}\label{sec:consistency}
\begin{figure}
  \centering 
\includegraphics[scale=0.6]{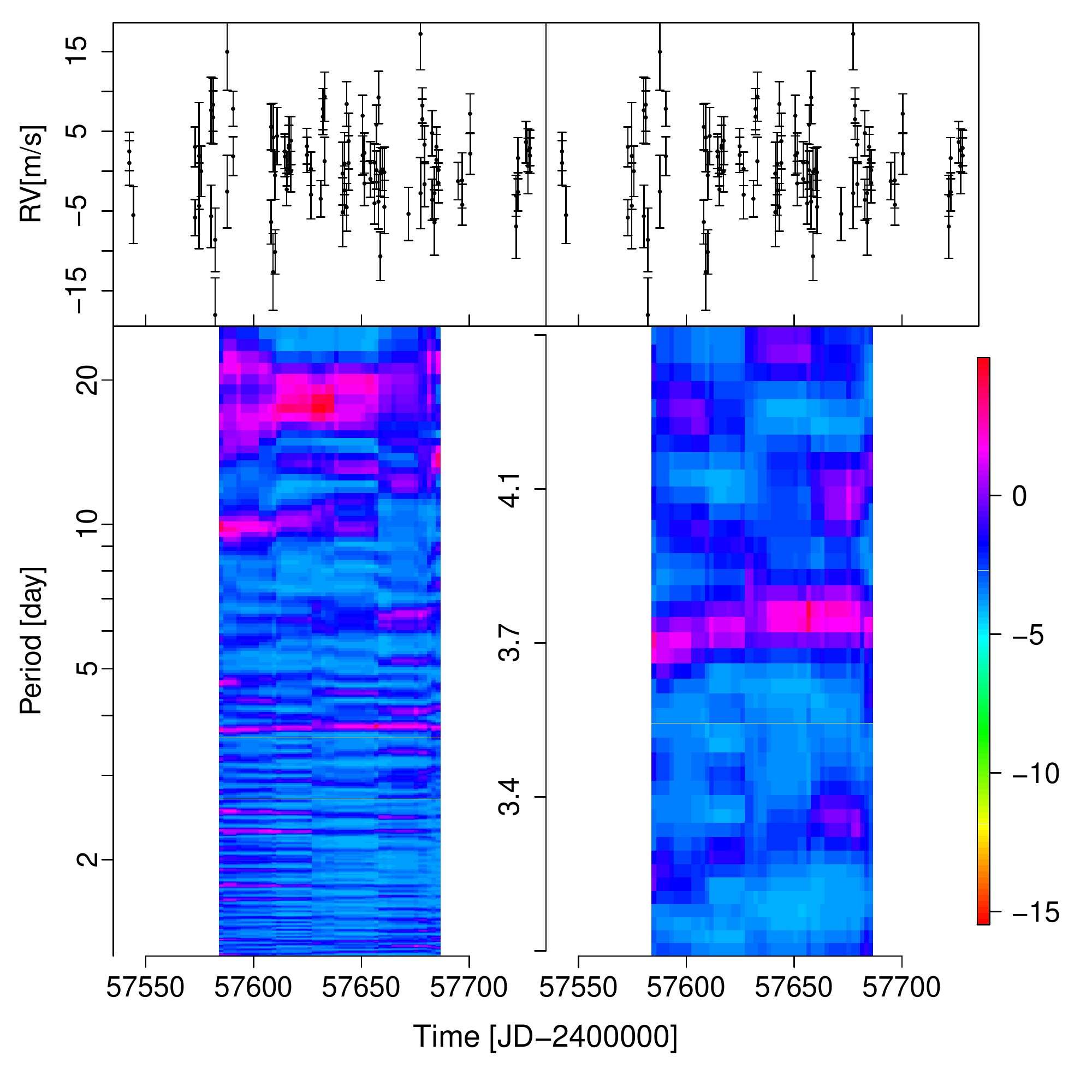}
\caption{The BFP-based moving periodogram for the second observation season of the HARPS data after subtracted by the best-fitting 2-planet model. The periodogram is shown in a long (bottom left) and short (bottom right) period range together with the data (upper panels). The color encodes the BF.}
\label{fig:mp}  
\end{figure}

We further investigate the consistency of the 3.8\,d signal over time using moving periodogram \citep{feng17a}, where BFP is calculated in a 100\,d moving window covering the data within 50 steps. Since there is a gap between the first observation season and the second one and there is much less data in the former, we show the moving periodogram for the second season in Fig. \ref{fig:mp}. We see consistent significance of the 3.8\,d signal in this observational season. An investigation of the first season also shows a signal around 3.8\,d. Such a time-invariant signal is unlikely to be caused by anything else but a planet. This warrants further investigations into the 3.8\,d signal through photometric follow up. 

\begin{figure}
  \centering
  \includegraphics[scale=0.6]{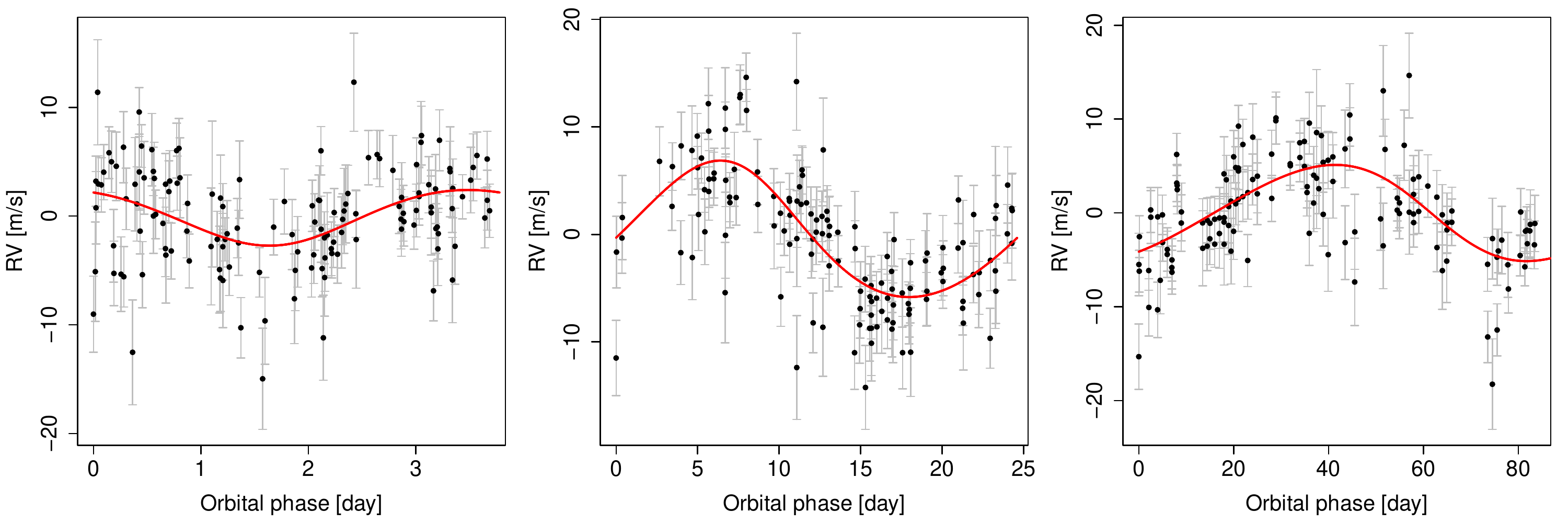}
  \caption{Phase folded radial velocity data and the radial velocity variation induced by LHS 1140 b (middle) and c (right) and d(left). In each panel, the best-fitting noise model and the other signal are removed from the data.}
\label{fig:phase}  
\end{figure}

Since the activity induced signal that is identified by D17 is not strong in the radial velocity data, it seems to be arbitrary to model the time-correlated noise using a Gaussian process with a strong prior determined by the rotation period. Moreover, such a process can not remove noise correlated in short time scales. With uninformative prior, Gaussian process tend to interpret signals as noise and lead to false negatives \citep{feng16}. Based on a comparison of different noise models in the Bayesian framework, we recommend MA(1) as the baseline model and the corresponding parameters as reference values. We show the phase curve for these three signals in Fig. \ref{fig:phase}. We report the physical parameters of these two candidates in Table \ref{tab:signal}. We use 1\% and 99\% quantiles to define the uncertainty intervals, which is appropriate because the posterior distribution of some parameters are not Gaussian. Considering the slight sensitivity of parameters to noise models, we find relatively larger uncertainty intervals than appropriate for reporting Keplerian signals.

\begin{table}
  \centering 
        \caption{The maximum {\it a posteriori} estimation of the parameters for three signals detected in the HARPS radial velocity data of LHS 1140. The uncertainties of parameters are represented by the values determined at 1\% and 99\% quantiles of the posterior densities. The noise models used to quantify signals are enclosed by square brackets in the column names. Since the 3.8\,d signal does not pass the BF threshold, we add a question symbol before LHS 1140 d. The mass is determined by adopting an inclination of 90 degree according to D17. }
\label{tab:signal}
%\footnotesize{
  \begin{tabular}  {l*{4}{c}}
\hline
&LHS 1140 b [white noise]&LHS 1140 b [GP] & LHS 1140 b [MA(1)] & LHS 1140 b [MA(2)]\\\hline
    $P$\,(d)&24.58 [24.45, 24.78]&24.55 [24.45, 24.85]&24.57 [24.43, 24.82]&24.60 [24.41, 24.83]\\
    $K$\,(m/s)&5.76 [4.63, 7.06]&5.66 [4.42, 7.18]&5.88 [4.34, 7.16]&5.40 [4.20, 7.28]\\
    $e$&0.11 [0.00, 0.23]&0.08 [0.00, 0.21]&0.10 [0.00, 0.23]&0.04 [0.00, 0.23]\\
    $\omega$\,(rad)&1.30 [0.09, 6.17]&1.67 [-3.03, 3.07]&1.29 [-3.07, 3.06]&1.77 [0.08, 6.19]\\
    $M_0$\,(rad)&3.34 [0.21, 6.09]&2.91 [0.24, 6.06]&3.31 [0.16, 6.15]&2.93 [0.13, 6.13]\\
$m$\,($M_\oplus$)&7.22 [5.20, 9.42]&7.12 [4.99, 9.50]&7.38 [4.95, 9.58]&6.81 [4.76, 9.53]\\
    $a$\,(au)&0.087 [0.078, 0.096]&0.087 [0.078, 0.096]&0.087 [0.078, 0.096]&0.087 [0.078, 0.096]\\\hline
&LHS 1140 c [white noise]&LHS 1140 c [GP] & LHS 1140 c [MA(1)] & LHS 1140 c [MA(2)]\\\hline
    $P$\,(d)&92.29 [90.27, 94.93]&93.07 [88.76, 96.05]&92.12 [89.39, 94.86]&92.31 [89.19, 95.14]\\
    $K$\,(m/s)&5.62 [4.04, 6.93]&5.64 [3.14, 7.07]&5.67 [3.80, 7.06]&4.92 [3.74, 7.14]\\
    $e$&0.00 [0.00, 0.20]&0.03 [0.00, 0.20]&0.03 [0.00, 0.20]&0.02 [0.00, 0.22]\\
    $\omega$\,(rad)&-3.08 [-3.06, 3.06]&-0.24 [-3.07, 3.08]&2.37 [-3.06, 3.05]&-3.13 [-3.08, 3.07]\\
    $M_0$\,(rad)&0.51 [0.07, 6.20]&4.14 [0.09, 6.22]&1.38 [0.08, 6.20]&0.59 [0.07, 6.21]\\
$m$\,($M_\oplus$)&11.01 [7.19, 14.38]&11.09 [5.63, 14.67]&11.11 [6.77, 14.60]&9.64 [6.58, 14.38]\\
    $a$\,(au)&0.210 [0.189, 0.232]&0.212 [0.190, 0.233]&0.210 [0.189, 0.232]&0.210 [0.189, 0.232]\\\hline
&?LHS 1140 d [white noise]&?LHS 1140 d [GP] & ?LHS 1140 d [MA(1)] & ?LHS 1140 d [MA(2)]\\\hline
    $P$\,(d)&3.78 [3.77, 3.79]&3.78 [3.72, 3.79]&3.78 [3.76, 3.79]&3.78 [3.77, 3.79]\\
    $K$\,(m/s)&2.36 [1.07, 3.71]&2.56 [0.64, 3.80]&2.70 [1.01, 3.80]&2.93 [1.12, 3.72]\\
    $e$&0.16 [0.00, 0.30]&0.07 [0.00, 0.27]&0.19 [0.01, 0.31]&0.15 [0.00, 0.30]\\
    $\omega$\,(rad)&2.53 [0.21, 6.03]&3.40 [0.20, 6.13]&3.82 [0.48, 5.44]&2.74 [0.14, 6.08]\\
    $M_0$\,(rad)&3.96 [0.16, 6.14]&3.34 [0.17, 6.10]&3.20 [0.71, 5.76]&3.73 [0.23, 6.05]\\
$m$\,($M_\oplus$)&1.57 [0.66, 2.52]&1.73 [0.38, 2.61]&1.79 [0.62, 2.61]&1.96 [0.69, 2.61]\\
    $a$\,(au)&0.025 [0.022, 0.028]&0.025 [0.022, 0.028]&0.025 [0.022, 0.028]&0.025 [0.022, 0.028]\\\hline
  \end{tabular}
\end{table}

\section{Conclusion}\label{sec:conclusion}
In summary, we find periodic signals of about 90\,d and 3.8\,d as planetary candidates. The former has been interpreted as an activity-induced signal by DT17. Our analysis shows that such a signal cannot be simply induced by stellar activity. Assuming an Earth-like density, we expect transits with depths of about 0.007 and 0.002 and durations of $\sim$1 and $\sim$3 hours for LHS 1140 c and d, respectively.  Although the transit depth of the 90\,d signal is larger than that of the 24.7\,d signal, the MEarth project may have missed this long period target due to the relatively short monitoring time span. The transit of LHS 1140 d is shallower than that of LHS 1140 b by a factor of 2.5 and thus may not be identified by the MEarth neural network as a signal for follow up observations. If confirmed by photometric observations, LHS 1140 c would be a mini-Neptune with a mass of 11\,$M_\oplus$ orbiting LHS 1140 at a nearly circular orbit with a semi-major axis of 0.21\,au. LHS 1140 d would be a Earth-sized planet located about 0.025\,au from LHS 1140. Our reanalysis of the radial velocity data provides a comprehensive strategy of disentangling signals from stellar activity and other correlated noise. 

\section*{Acknowledgements}
This work is supported by the Science and Technology Facilities Council [ST/M001008/1].

%%%%%%%%%%%%%%%%%%%%%%%%%%%%%%%%%%%%%%%%%%%%%%%%%%%%%%%%%%%%%%%%%%%%%%%%%%%%
%\section*{Acknowledgements}
%FF, MT and HJ are supported by the Leverhulme Trust (RPG-2014-281) and the Science and Technology Facilities Council (ST/M001008/1). We used the ESO Science Archive Facility to collect radial velocity data sets. 
%\section*{Method}

\bibliographystyle{aasjournal}
\bibliography{nm}  
\end{document}